\documentstyle[prl,aps,epsf]{revtex}
\draft
\begin{document}
\twocolumn[\hsize\textwidth\columnwidth\hsize\csname @twocolumnfalse\endcsname

\title{Crossover from Intermittent to Continuum Dynamics for Locally
Driven Colloids} 
\author{C. Reichhardt and C.J. Olson Reichhardt} 
\address{ 
Theoretical Division and Center for Nonlinear Studies,
Los Alamos National Laboratory, Los Alamos, New Mexico 87545}

\date{\today}
\maketitle
\begin{abstract}
We simulate a colloid with 
charge $q_d$ driven through a disordered 
assembly of interacting colloids with charge $q$
and show that, for $q_d\approx q$, the velocity-force 
relation is nonlinear and the velocity fluctuations of the driven
particle are highly intermittent with
a $1/f$ characteristic. When $q_d\gg q$, the
average velocity drops, the velocity
force relation becomes linear, and the velocity 
fluctuations are Gaussian. 
We discuss the results in terms of a crossover
from strongly intermittent heterogeneous dynamics to continuum dynamics.
We also make several predictions for the transient response in
the different regimes. 
\end{abstract}
\vspace{-0.1in}
\pacs{PACS numbers:82.70.Dd}
\vspace{-0.3in}

\vskip2pc]
\narrowtext
An individual particle driven through an overdamped medium 
exhibits a linear
velocity vs applied force relation.
When quenched disorder is added to the system,
a critical threshold driving force must be applied
before the particle can move, and once motion begins, the velocity-force
curves can be nonlinear.
Examples of overdamped systems with quenched disorder
that exhibit threshold forces and nonlinear velocity force
curves include driven vortices
in type-II superconductors or Josephson junctions 
\cite{Dominguez}, 
sliding charge density waves \cite{Fisher,Higgins}, driven magnetic 
bubble arrays \cite{Westervelt}, and charge transport in 
assemblies of metallic dots \cite{Middleton}. These systems
have been studied extensively and exhibit 
a variety of dynamic phases as well as power law velocity force curves. 

Another system that has been far less studied 
is an overdamped particle moving in 
the {\it absence} of quenched disorder but
in the presence of a disordered two-component background of 
non-driven particles.
Since there is no quenched disorder, 
simple overdamped motion with an increased damping constant might be expected.
Instead,
a critical threshold force $F_{th}$ for motion exists
and the velocity force curves 
are nonlinear with the power law form
$V = (F -F_{th})^{\beta}$, as shown in  
recent simulations \cite{Hastings} and experiments \cite{Weeks} 
for individual colloids driven through a background of
non-driven colloids.
Unlike systems with 
quenched disorder, for $F<F_{th}$ the entire system must move along with 
the driven particle.  For $F>F_{th}$ the driven particle is able to shear
past its neighbors.
The simulations give $\beta=1.5$ for colloids interacting with
a screened Coulomb potential.
The experiments were performed with
lightly charged colloids where steric interactions are important,
and give $\beta\approx 1.5$ 
in the lower density limit 
\cite{Weeks}.
A particle moving 
through a viscous fluid can be regarded as interacting with many much
smaller particles.  In this limit,
the surrounding medium can be replaced with a continuum and the velocity force
curves are linear.  In the case 
of a single driven colloid,
it is not known when and how the dynamics change from nonlinear 
to linear, since it is
expected that 
the system passes to the continuum limit when    
the colloids in the surrounding medium
are small.  This change could be a sharp transition, or it could occur as
a cutoff of the scaling, or it could occur through a continuously changing
exponent.

In this work we consider a single colloid 
of charge $q_d$ driven through a $T=0$
disordered two-component assembly of other colloids with average charge $q$. 
We find that when 
$q \approx q_d$,
the velocity force curves have a power law form with $\beta > 1$ 
that is robust over two decades and for
different system sizes. 
In this regime the velocity fluctuations 
of the driven colloid
are highly 
intermittent and 
the colloid velocity $V$ frequently drops nearly to zero 
when background colloids trap the driven colloid until rearrangements
release it
and $V$ jumps back to a higher value.
The fluctuations in $V$ have a highly skewed distribution
and $1/f$ noise fluctuation properties. 
As $q_d$ increases for fixed drive $F_d$,
the average velocity $V$ drops,
the velocity fluctuations become Gaussian, and $\beta$ is reduced. 
When $q_d\gg q$, the driven colloid
interacts with a large number of surrounding colloids and
forms a circular depletion region, while
for $q_d\ll q$, 
the background colloids act as stationary disorder
and the velocity force curves are linear. 
Our system can be experimentally realized for dielectric colloids driven
with optical traps or magnetic colloids driven with external 
magnetic fields.
In systems where it is difficult to vary the charge on individual particles, 
$q_d$ could be increased by capturing 
a large number of 
particles in a single optical trap and
dragging the assembly through the background. 
Other related systems include
dragging different sized particles through granular media
\cite{Albert}.  
There have been several proposals to use 
individual particle manipulation as a new microrheology method for
examining frequency responses in soft 
matter systems \cite{Levine}. It would 
be valuable to understand
under what conditions the individual particle is in 
the continuum overdamped regime or in the nonlinear 
regime in the dc limit. 

We consider a substrate-free, zero temperature, two-dimensional 
system with periodic boundary conditions in the $x$ and
$y$ directions. 
A binary mixture of $N=N_c-1$ background colloids, charged
with a ratio $q_1/q_2=1/2$, 
interact with a repulsive screened Yukawa potential,
$V(r_{ij}) = (q_{i}q_{j}/|{\bf r}_{i} - {\bf r}_{j}|)
\exp(-\kappa|{\bf r}_{i} - {\bf r}_{j}|)$, 
where
$q_{i(j)}$ is the charge and ${\bf r}_{i(j)}$ is
the position of colloid $i$($j$) and
$1/\kappa$ 

\begin{figure}
\center{
\epsfxsize=3.4in
\epsfbox{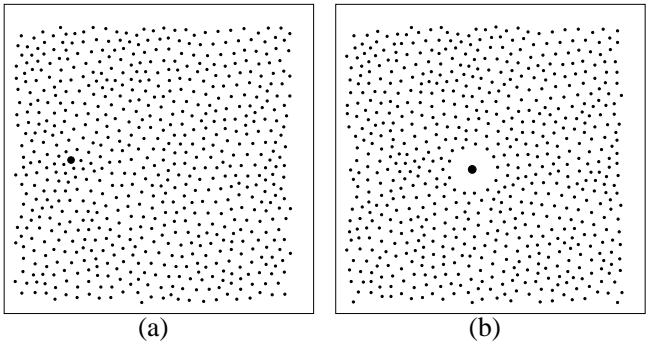}}
\caption{Small dots: background of binary charged colloids.  Large dot:
Driven colloid with charge
(a) $q_{d}/q = 1.33$ and
(b) $q_{d}/q = 67$.
}
\end{figure}

\noindent
is the screening length which is 
set to $2$ in all our simulations.
Throughout the paper we refer to the average background charge
$q=(q_1+q_2)/2$.
The initial disordered 
configuration for the two-component background of colloids 
is obtained by annealing 
from 
a high temperature. 
An additional driven colloid 
with charge $q_d$
is placed in the system and a constant 
driving 
force ${\bf F}_{d}=F_d{\hat{\bf x}}$ is applied only to that colloid.
The overdamped equation of 
motion for
colloid $i$ is 
\begin{equation}
\eta\frac{d {\bf r}_{i}}{dt} = {\bf F}_{i}^{cc} + {\bf F}_{d} + {\bf F}_{T}
\end{equation}
where 
${\bf F}_{i}^{cc} = -\sum_{j \neq i}^{N_{c}}\nabla_i V(r_{ij})$,
$\eta=1$, 
and 
the thermal force ${\bf F}_{T}$ comes 
from random Langevin kicks. We have considered various temperatures
and discuss the $T=0$ case here. 
We have previously used similar  
Langevin dynamics for colloids under nonequilibrium
and equilibrium conditions \cite{ourstuff}.   
The interaction range is assumed much larger
than the physical particle size, and in this low volume
fraction limit, hydrodynamic interactions can be neglected and may be 
strongly screened \cite{Riese}.
To generate velocity-force curves, 
we set $F_{d}$ to a fixed value 
and measure the average velocity  of the driven colloid $<V>$ in the
direction of the drive for several million 
time steps to ensure that a steady state is reached.  The drive is
then increased and the procedure repeated. 
Near the depinning threshold $F_d \gtrsim F_{th}$, the relative 
velocity fluctuations are strongly enhanced.
In the absence of any other particles, the driven colloid moves at 
the velocity of the applied drive giving a linear velocity force
curve.    
In this work we consider system 
sizes of $L = 24$, 36, and 48 with a fixed colloid density of $1.1$.   
This is four times denser than the system considered in Ref.~\cite{Hastings}.

In Fig.~1 we show images from the two limits of our system. In Fig.~1(a)
the driven colloid (large black dot) has $q_{d}/q = 1.33$ and
is similar in charge to the colloids forming
the surrounding disordered medium (smaller black dots).  Fig.~1(b) 
illustrates the case 
$q_{d}/q = 67$, where a large depletion zone forms around the driven colloid.

To illustrate the scaling in the velocity force curves, 
we plot representative $V$ vs $F_{d} - F_{th}$ curves in Fig.~2
for 

\begin{figure}
\center{
\epsfxsize=3.2in
\epsfbox{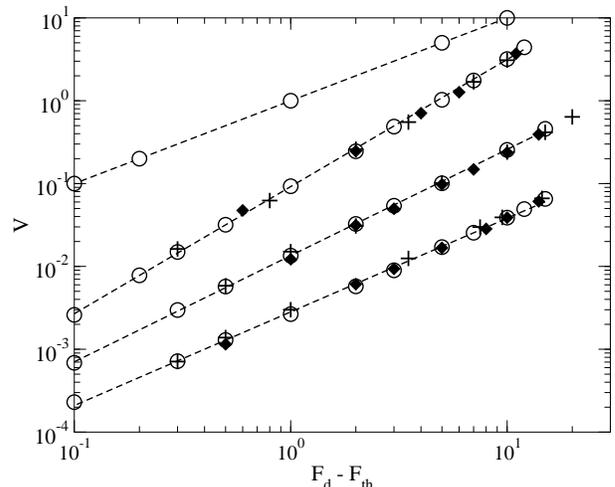}}
\caption{
Velocity $V$ vs applied force $F_{d} - F_{th}$, where $F_{th}$ is the
threshold force for motion.  From top to bottom, $q_{d}/q = 0.25$,
1.33, 13, and $67$. Dotted lines: fits with, from top to bottom,
$\beta = 1.0$, 1.54, 1.28, and 1.13.  System 
sizes: $L = 24$ (open circles), $36$ (plus signs), and $48$ (filled
diamonds).       
}
\end{figure}

\noindent
varied $q_{d}/q$ and different system sizes. 
Here $F_{th}$ is the threshold velocity and 
the charge of the driven colloid 
increases from the top curve
to the bottom.
All of the curves
have a power law velocity force scaling of the form 
\begin{equation}
V \propto (F_{d} - F_{th})^{\beta} 
\end{equation} 
This scaling is robust over two decades in driving force. 
To test for finite size effects, we conducted simulations with 
systems of size $L = 24$, $L = 36$,
and $L = 48$, indicated by different symbols in Fig.~2, 
and we find that the same scaling holds for all the 
system sizes. 
The scaling of the velocity force curves for the small charge
$q_{d}/q = 0.25$ is linear, as seen in previous
simulations \cite{Hastings}. In this regime the driven colloid does 
not cause any distortions
in the surrounding media as it moves. 
This situation is very similar to a 
single particle moving in a quenched background, where it is known 
that the velocity force curves scale 
linearly or sublinearly \cite{Fisher}.
As the charge of the driven colloid increases, 
the scaling exponent initially rises, as shown for the 
case of $q_{d}/q = 1.33$ with $\beta = 1.54$, 
but the exponent decreases again for the
more highly charged driven colloids, since 
$q_{d}/q = 13$ gives $\beta = 1.28$
and $q_{d}/q = 67$ gives $\beta = 1.13$. 
As $q_{d}/q$ increases, the average
velocity at fixed $F_d-F_{th}$ decreases
when more background colloids become involved in the motion.  

We plot the changes in the scaling exponent $\beta$ with varying
$q/q_{d}$ from a series of simulations in Fig.~3, which
shows three regions.   
For low $q_{d}/q$, the motion
is mainly elastic with $\beta$ near 1.  As $q_d/q$ approaches 1,
the driven colloid charge becomes of the same order as that of the
surrounding medium and
the motion becomes plastic 
with 
$\beta \approx 1.5$. As $q_d/q$ increases further, 
$\beta$ decreases 
approximately logarithmically
toward 1, indicating that 
the motion 
of the surrounding medium is becoming more continuum-like.  
The maximum value of 

\begin{figure}
\center{
\epsfxsize=3.2in
\epsfbox{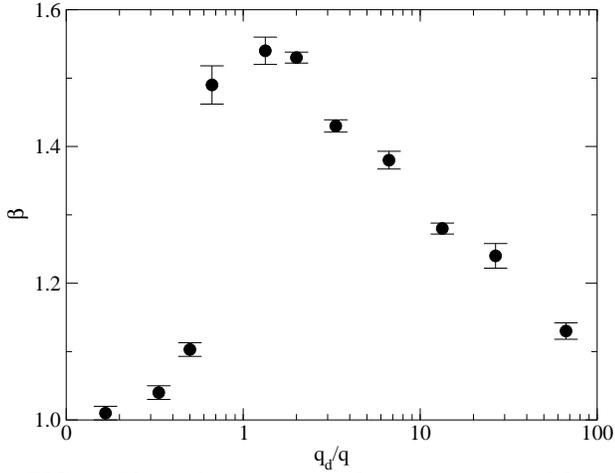}}
\caption{ The scaling exponent $\beta$ vs $q_{d}/q$ extracted from the  
velocity force curves.   
}
\end{figure}

\noindent
$\beta$ falls at higher $q_d/q$ for less dense systems,
such as that in Ref.~\cite{Hastings}, and at lower $q_d/q$ for denser
systems.

For driven colloids with $q_d/q\ll 1$, the background colloids act as
a stationary disorder potential.  The driven 
colloid passes through
this potential, deviating around 
background colloids as necessary, but
the background colloids do not respond to the presence of the driven 
particle and remain essentially fixed in their locations.  In contrast, when 
$q_d/q\gg 1$, the driven colloid strongly distorts the background 
and forms a depletion zone which moves with the driven colloid.
Thus a large number of background colloids must rearrange in order to
allow the driven colloid to pass, producing a continuum-like behavior.
Between these two limits, when $q_d/q \approx 1$, 
no depletion zone forms, but when the driven colloid moves, it distorts
the background which deforms plastically in order to allow the
driven colloid to pass.  Here we find intermittent motion in which
the driven colloid sometimes slips past a background colloid similar
to the $q_d/q\ll 1$ case, but at other becomes
trapped behind a background colloid and pushes it over
some distance, similar to the $q_d/q\gg 1$ case.  It is in the regime of
this complex motion, when all of the charges are similar in magnitude,
that the strongest deviation from linear response, $\beta \approx 1.5$,
occurs.

We next consider the velocity fluctuations of the driven colloid in
the regime where $\beta \approx 1.5$ as well as 
in the high $q_d/q$ regime where
$\beta$ starts to approach 1. In Fig.~4(a) we plot a segment
of the time series of the instantaneous driven colloid velocity  
for the case of $q_{d}/q = 1.33$ at a drive producing an average 
velocity of $V = 0.0425$.  The motion is highly intermittent and at
times the colloid temporarily stops moving in the direction 
of the drive.
When the driven colloid is trapped, strain accumulates 
in the  
surrounding media until
one or more of the surrounding colloids suddenly 
shifts by a distance larger than the 
average interparticle spacing 
and the driven particle begins to
move again.
As the drive is further increased, 
the length of the time intervals during which the driven 

\begin{figure}
\center{
\epsfxsize=3.5in
\epsfbox{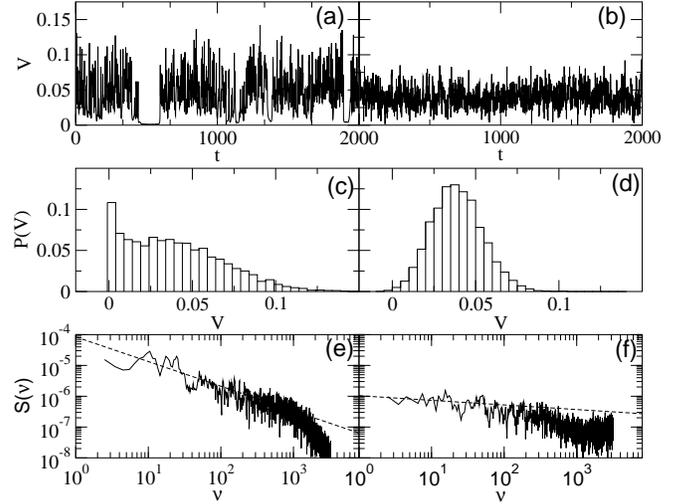}}
\caption{
(a) A segment of the 
instantaneous driven particle velocity time series $V(t)$
for a system with $q_{d}/q = 1.33$ at
a drive giving an average velocity of $V=0.0425$. 
(b) A segment of $V(t)$ for $q_d/q=67$ for a drive giving the same
average velocity as in (a).
(c) The velocity distribution $P(V)$ for the system shown in (a). 
(d) $P(V)$ for the system in (b). 
(e) Power spectrum $S(\nu)$ of
the time series in (a). Dashed line: power law fit 
with exponent $\alpha = -0.8$.
(f) $S(\nu)$ of the time series in (b).
Dashed line: power law fit with $\alpha = -0.15$.  
}
\end{figure}

\noindent
colloid
is stopped decreases.

In Fig.~4(b) we show 
$V(t)$
for a system with $q_{d}/q = 67$ 
where the applied force gives the same average velocity
as in Fig.~4(a). Here the amplitude of the 
velocity fluctuations is much smaller than the $q_d/q=1.33$ case
and there are no intermittent stall periods. 
This strongly charged driven colloid
is interacting with a much larger number of surrounding
colloids than the weakly charged driven colloid would, and  
as a result, it cannot be trapped behind a
single background colloid, giving much smoother motion.
We note that we find no intermittent behavior for the strongly
charged driven colloid even at the lowest applied forces.
We also measured the variance of the transverse velocity fluctuations
$V_y$, and find that it
decreases with increasing $q_d/q$ roughly as a power law with an exponent of 
-1.7. This decrease occurs since a larger number of
background colloids are contributing to the fluctuations experienced
by the driven particle, leading to a smoother signal.

In Fig.~4(c) we plot the histogram of the velocity fluctuations
$P(V)$ for the time series shown in Fig.~4(a).  The fluctuations 
are non-Gaussian and are heavily
skewed toward the positive velocities 
with a spike at $V = 0$ 
due 
to the intermittency. 
We note that in simulations 
of vortex systems
where nonlinear velocity force scaling 
occurs,
bimodal velocity distributions are also observed
when
individual vortices are intermittently pinned for
a period of time before moving again \cite{Marchetti}. 
Non-Gaussian velocity 
fluctuations are also found in
sheared dusty plasmas \cite{Woon}.

\begin{figure}
\center{
\epsfxsize=3.4in
\epsfbox{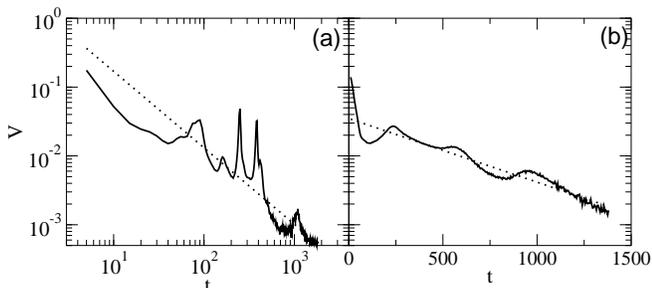}}
\caption{
The transient velocity response to a sudden applied force 
of $F_{d}/F_{th} = 0.8$. (a) $q_{d}/q = 1.33$; the dashed line is 
a power law fit with an exponent of $\alpha = -1.1$. 
(b) $q_{d}/q = 67$; the dashed line is an exponential fit. 
}
\end{figure}

\noindent
For higher drives, we find that
the average velocity increases; however,  
the histograms remain highly skewed 
for all charge and drive regimes
where the scaling in the 
velocity force curves give a large $\beta \sim 1.5$.
For comparison, in Fig.~4(d) 
we plot $P(V)$ for the large $q_d/q$ system shown in Fig.~4(b). Here
the histogram has very little skewness and fits well to a Gaussian 
distribution. For other drives 
at this charge ratio we observe similar Gaussian distributions
of the velocity.
In general, we find decreasing skewness in the velocity distributions
as $q_{d}/q$ increases.  The Gaussian nature of the fluctuations 
is also consistent with the interpretation that, as $q_d/q$ becomes large, 
the system enters the continuum limit.  

In Fig.~4(e), we show that the power spectrum
$S(\nu )$ for the time series in Fig.~4(a) has 
a $1/f^{\alpha}$ form with $\alpha = 0.8$. 
Throughout the $\beta\approx 1.5$
regime we find similar spectra with $\alpha = 0.5$ to $1.1$,
indicative of intermittent dynamics.
For comparison, in Fig.~4(f) we show that $S(\nu )$ for the high $q_d/q$
case has a white velocity spectrum characteristic,
indicative of the absence of long time correlations in the velocity.
In general, $\alpha$ decreases with increasing $q_{d}/q$.  
For the small charge 
regime of $q_{d}/q\ll 1$, where the driven colloid moves without
distorting the background, we also observe a white noise spectrum.
Here the velocity fluctuations are determined 
by the static configuration
of the background particles, and $P(V)$ does not show any 
intermittent periods of zero
velocity because if the colloid becomes trapped in this regime, 
there can be no rearrangements of the surrounding medium to untrap it.

To explore the transient behavior of the system,  
we consider the effect of a suddenly applied subthreshold drive of
$F_{d}/F_{th} = 0.8$. 
In Fig.~5(a) we show the transient velocity response for a system with
$q_{d}/q = 1.33$. Here the velocity relaxation 
is consistent with a power law decay, 
$V(t) \propto t^{-1.1}$.
The driven colloid translates by several lattice constants before 
coming to rest with respect to the surrounding medium.
The large velocity oscillations that appear at longer times are 
due to the local plastic rearrangements 
of the surrounding colloids as the driven colloid passes. 
We find a power law decay in the transient response
for values of $q_{d}/q$ that give $\beta > 1.28$. 
If the suddenly applied subthreshold drive is small enough that
no local rearrangements of the surrounding medium are possible,
then an exponential decay of the velocity occurs instead.
In Fig.~5(b) we show the transient response for the case of
$q_{d}/q = 67$. Here the relaxation is fit to 
$V(t) \propto \exp(-t)$. The large velocity fluctuations 
that appeared for the smaller $q_{d}$ are absent. For all the large values of 
$q_{d}/q$ we observe an exponential velocity relaxation, and
we also find exponential relaxation
for very small charges $q_{d}/q \ll 1.0$.

In summary, we have studied a single
colloid with varying charge 
driven through a disordered background of other
colloids in the absence of quenched disorder.
When the charge of the
driven colloid is close to the same as that of the 
surrounding colloids, a nonlinear
power law velocity force curve appears with an exponent near $\beta=1.5$. 
In this regime, the time dependent velocity is intermittent with 
a highly skewed velocity distribution and $1/f^{\alpha}$ noise fluctuations.   
As the charge of the driven colloid is increased, 
the velocity force characteristic becomes more linear while
the effective damping from the background colloids increases. 
The number of colloids that
interact with the driven colloid increases and 
the velocity fluctuations become 
Gaussian with a white noise spectrum. We predict that 
in the nonlinear regime, the transient velocity responses are
of a power law form, 
while in the linear regime the transient velocity is exponentially
damped. We interpret our results as a crossover 
in the response of the background colloids from intermittent dynamics
when the driven and background colloids are similarly charged
to continuum dynamics for a highly charged driven colloid.

We thank E.~Weeks and M.B.~Hastings 
for useful discussions. This work was supported by the U.S.~Department
of Energy under Contract No.~W-7405-ENG-36.

\end{document}